# An assistive HCI system based on block scanning objects using eye blinks

Supriya Sarker
Dept. of Computer Science & Engineering
Chittagong University of Engineering & Technology
Cittagong, Bangladesh
sarkersupriya7@gmail.com

Md. Shahraduan Mazumder
Dept. of Computer Science & Engineering
World University of Bangladesh
Dhaka, Bangladesh
mjmajumder55@gmail.com

Md. Sajedur Rahman
Dept. of Electrical and Electronic & Engineering
World University of Bangladesh
Dhaka, Bangladesh
sajid.peec.wub@gmail.com

Md. Anayt Rabbi
Dept. of Computer Science & Engineering
World University of Bangladesh
Dhaka, Bangladesh
anaytrabby@gmail.com

*Abstract*—**Human-Computer Interaction (HCI) provides a new communication channel between human and the computer. We develop an assistive system based on block scanning techniques using eye blinks that presents a hands-free interface between human and computer for people with motor impairments. The developed system has been tested by 12 users who performed 10 common in computer tasks using eye blinks with scanning time 1.0 second. The performance of the proposed system has been evaluated by selection time, selection accuracy, false alarm rate and average success rate. The success rate has found 98.1%.**

## I. INTRODUCTION

In today's world communication is the most vital subject for a human to remain in touch with the environment [1]. During the era of digital technology, there is no alternative to a computer for communication. Computer with a couple of installed software in it can allow us contributing to our social lives, make our learning process smarter, and grow us as an entrepreneur. But it is likely possible when a person capable of using a computer by his or her own. A person with any kind of disability is more likely to experience adverse conditions both socially and economically, for example, lack of education, health, employment, and higher poverty [2]. People are so busy with their own lives that it becomes pretty difficult to spend time for others, even for disabled or sick family members. People suffering from motor impairments, spinal muscular atrophy (SMA), and Amyotrophic Lateral Sclerosis (ALS) or spinal cord damages face complicacy to live a sound life because of their inability [3]. Therefore, an effective means of communication without verbal communication and body muscles movements is necessary to improve the facilities in the lives without care takers. User's eye blink has been used to control the computer [4]. A brief description of the existing approaches is presented in Section 2. The methodology of the proposed including eye detection, monitor object scanning, is described in Section 3. The experimental results are presented in Section 4. Finally, Section 5 concludes the paper.

## II. EXISTING APPROACHES

Nowadays Human-Computer Interaction (HCI) has been received a great deal of attention in the area of assertive technology. From the literature review, it is clearly perceivable that enormous amount of eye-tracking based computer-controlled system has been developed for HCI. To develop neural communication with computers various biomedical signals have been used such as Electrooculogram (EOG), Electro-Encephalogram (EEG), Electromyogram (EMG) [5]. In [6] a system has been implemented a head operated Human-Computer Interaction system where mouse pointer on the computer monitor screen has followed the direction of user's head tilt and activation of click has been done by eyebrow movements detected by the accelerometer-based tilt sensor. The system also introduced a voice-recognition system to recognize small letters pronounced by a disabled person. However, it was quite uncomfortable for users to wear the headset device all the time and chance of cause issues in the muscles of shoulder. In [7] a system has been proposed where a Bluetooth headset with one sensor on the forehead and the other three sensors on the left ear of the user has placed to acquire pseudo electromyography (EMG) signal of user's blink of eyes. The cognizant blink generated a variation in EMG signal on the user's forehead which eventually is detected and filtered by their proposed algorithm to trigger the scanning process of virtual keyboard. However, EMG has noisy characteristics that are caused by inherent equipment noise, electromagnetic radiation, motion artifacts, and the interaction of different tissues. It makes difficulties in analyzing EMG signal [5]. In [8, 9, 10] the authors proposed image processing method combining Haar-cascade classifier (HCC) to detect face and template matching for eye and eye-blink tracking. In [9] an alphabet tree was built from where patients can choose words by blinking left or right eyes to make a sentence.

Hence, the major contribution made in this research is to develop a lightweight, easy to use, harmless assistive

Human-Computer Interaction system based on block scanning techniques using eye blinks.

### III. PROPOSED METHODOLOGY

This section explains the system architecture including layout of human eyes, the process of transmission of analog signal generated from eye blink, and block scanning mechanism.

#### A. System Architecture

In the proposed system, an IR reflectance sensor (QRE1113) mounted on the eyeglass frame of the user. Fig 1 illustrates the block diagram of the proposed system. When user blinks eye it receives a signal. We use two Bluetooth modules (HC 05) - the first one is used as master mode and the second one is used in slave mode. For better understanding, we called the Bluetooth in master mode is as Bluetooth1 and the Bluetooth in slave mode is as Bluetooth2. The signal received by the sensor is being transmitted to the microcontroller of Arduino Nano and the microcontroller transfers the signal to the Bluetooth1. Bluetooth1 sends the signal to Bluetooth2 which is being transferred to the microcontroller of Arduino Pro-micro. The Analog to Digital Conversion system (ADC) of Arduino Pro-micro converts the analog signal to digital data. With the help of the USB Human Interface Device system (HID) of the microcontroller of Arduino Pro-micro the digital data moves to the user's computer and controls the computer according to the user's blinking command.

#### B. Layout of Human Eye

The layout and functions of the human eyes are very complicated. Each eye continually adapts the amount of reflected light, focuses on objects from different distances, and produces continuous images that are transmitted to the brain every moment [11]. Our eyes are a somewhat irregular-shaped globe, approximately an inch in diameter. The structure of eyes includes Iris, Cornea, Pupil and Sclera, Retina. The main parts are described below:

- Iris is the colored area of the eye and encircles the pupil. This portion manages light regularly, accommodates with the change of light. It works like a camera.
- Cornea is a clear, dome-like layer that wraps the pupil, iris, and interior parts or fluid-filled area between the cornea and the iris. The cornea must remain clear to reflect light.
- Pupil is the black circular opening in the iris that lets light in and sclera is the white area of our eyes.
- Lens lies behind the iris and pupil. Light projects through our pupil and lens to the back of the eye. retina contains special light-sensing cells covered inside lining of the eye. It converts light into electrical impulses. Optic nerve carries converted impulses to the brain. Light reflects off an object, and if that object is in our field of vision, it enters the eyes.

#### C. Transfer of Eye Blinks

The proposed system is divided into two sections. The first section comprised of Arduino Nano, IR reflectance sensor, Bluetooth, power source and is placed in the glass frame of the user. The situation when user either blinks eyes or stares at the monitor is captured by IR sensor as a change of voltage level. At the time, when user blinks an eye no light transfer to the IR sensor and a rapid drop in voltage encounters. This blink point is considered as a click on the on monitor screen of the user computer. The change of voltage is clearly shown in Fig. 2.

The second section consists of Bluetooth, 10k potentiometer, Arduino Pro-micro that is connected with the user's PC through the wire. IR sensor is comprised of two parts- an IR emitting LED and an IR sensitive phototransistor. The battery attached with master mode so that the device can be portable.

From the IR emitting LED, light transmits to the eyes when they are at open state and reflects on the surface of the eye pupil. The reflected light transfers to the photodiode of IR sensor. The IR photodiode is sensitive to the IR light emitted by an IR LED. The photodiode resistance and output voltage change in proportion to the IR light received. Then photodiode sends it as an analog signal to the microcontroller of Arduino Nano. The microcontroller sends the signal to Bluetooth in master mode. After that, signal transfers to Bluetooth in slave mode. Bluetooth in slave mode sends it to Arduino Pro Micro. Arduino Pro-micro has HID (Human Interface Device) and ADC (Analog to Digital Conversion) system so that it can convert an analog signal into a digital signal. Thus, the open and blinking state of the user eye converted into digital information and being used like controlling mouse pointer on the on screen computer monitor.

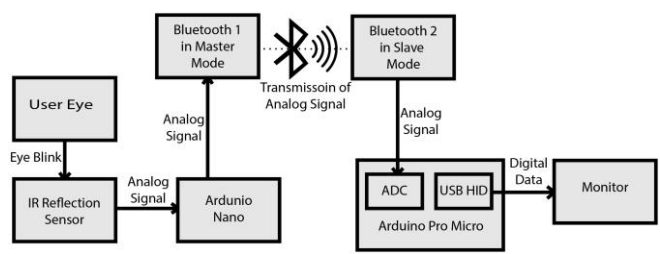

Fig. 1. Block diagram of the proposed system

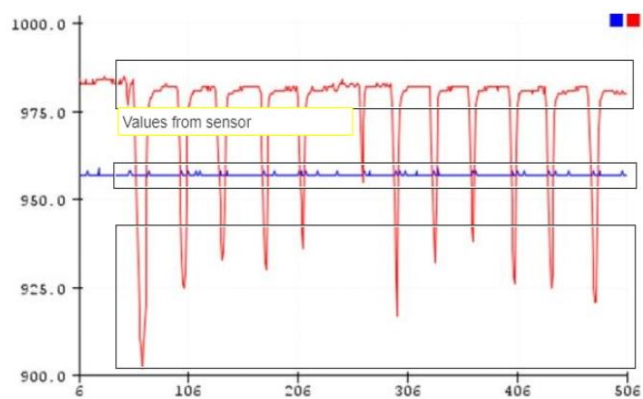

Fig. 2. Transmission of analog signal from phototransistor

## D. Transmission of Analog Signals

The analog signal with different voltage value continuously transmitted from phototransistor to the microcontroller of Arduino Nano. We plotted the transmitted analog signal using serial plotter that is shown in Fig. 2. In Fig. 2, we plotted voltage frequency in Y-axis and time in X-axis. We categorize the voltage of the analog signal in three classes, they are- base voltage, garbage voltage, and threshold voltage.

When the user's eyes remain in static and motionless, there is a very little change of the voltage of the analog signal. This portion of the analog signal is named as base voltage. Threshold voltage needs to adjust according to the user's voluntary blinks. The sensor always senses frequencies. The lower voltage than the threshold voltage is considered as voluntary blink. According to user analyses, 87% of them are using the device with this threshold value and considered as a voluntary blink. The involuntary blinks of users are considered as a garbage voltage value. The threshold voltage is stored in a variable resistor. The variable register facilitates users by modifying the threshold voltage. The analog signal that is being converted to digital signal by Arduino Pro-micro transfers to the potentiometer. Then potentiometer receives the signal and calibrates it to fix the point of blink on the screen. Finally, the objects on the computer monitor is being scanned and the desired point are being clicked by the mouse pointer.

## E. Block Scanning Mechanism

Scanning is a method of consecutively emphasize an item on a computer screen and trigger a switch when item is being emphasized. As Graphical User Interface (GUI) is getting very popular, navigation to random locations on the computer screen has become more important in assistive technology. Cartesian scanning and polar scanning are two common scanning mechanisms. In Cartesian scanning, the cursor moves progressively in a parallel direction to the edges of the monitor screen. In Polar scanning, a direction is selected before moving along a fixed bearing. But the cartesian method was shown to be faster and most effective scanning strategy over the entire screen [12].

In the block scanning techniques, the monitor screen area is repetitively subdivided into equally-sized sub-areas that are shown in Fig. 3, 4, 5, 6 for four iterations. The user has to select a sub-area that contains the intended target when program control switched to one sub-area to another. The segmentation process continues a certain number of repetitions and eight-directional scanning is activated in the chosen sub-area. The eight directions are up, up-left, left, left-down, down, downright, right, right-up**.** Block scanning is a better technique than that of others because it divides the screen into four equal-sized partitions for four iterations and then switches to eight-directional scanning [13]**.** A sample monitor screen scanning is demonstrated in [14]. In [15] the performance of block scanning techniques proved better up to four iterations than other methods.

## IV. RESULTS AND ANALYSIS

This section demonstrates the real-time monitor scanning process applying block scanning techniques along with system performance and users' performance while using the system.

### A. Monitor Scanning Process

The process of monitor object scanning is shown in fig 7, 8, 9, and 10. At first step, the monitor is being scanned (Fig. 7), then the repeatedly subdivided into subareas (Fig. 8, 9). At the last step an object on the monitor screen has been selected (Fig. 9) and option to perform are shown (Fig. 10). Fig. 11 demonstrates the typing process using the on-screen keyboard.

### B. Experimental Results

Twelve users (6 male, 6 female) ranging in age from 13 to 45 spontaneously have participated in the evaluation of the proposed system.

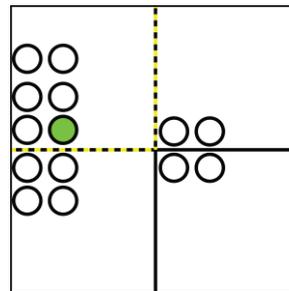
Fig. 3. Monitor has segmented into four sub-areas

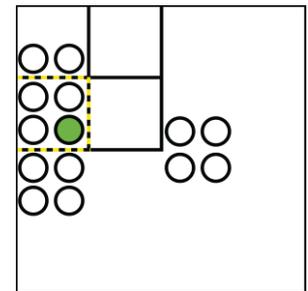
Fig. 4. Desired part has selected and divided into two parts

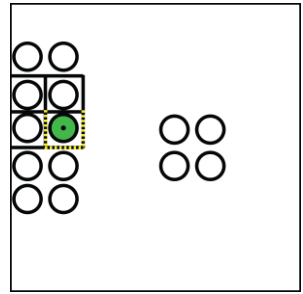
Fig. 5. One of the selected part has divided into four parts

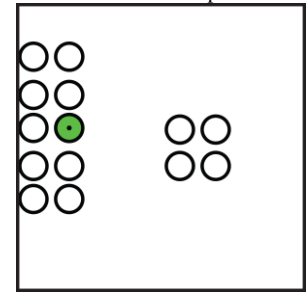
Fig. 6. Desired option selected

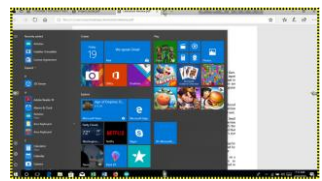
Fig. 7. Full monitor scanning process

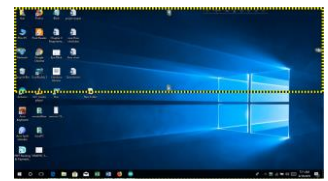
Fig. 8. Segmentation of screen into sub-areas

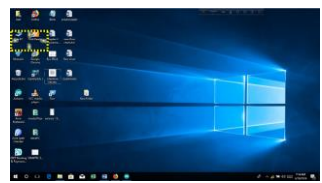
Fig. 9. After third step of segmentation target object is identified

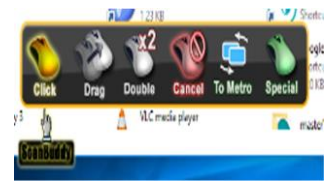
Fig. 10. Task selection

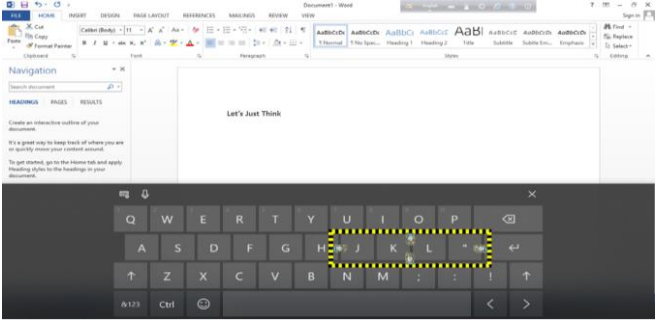

Fig. 11. Typing process using onscreen keyboard

No one had experience of using eye-controlled HCI systems beforehand. The system is not trained but the users need to practice to operate the system using blinks effectively. The blink is the regular voluntary blink, so when the monitor is being scanned users need to blink eye to select the desired block. So, there is no predefined blink rate or duration of blinks. The only variation is threshold value need to adjust according to voluntary blinks of the user's eye. It controls the rate of considerable blinks i.e, clicks on the monitor while scanning. Before starting the actual experiment and performing any measurement, a training session was conducted for all of the users. We have selected 10 tasks to perform by the users. The tasks are listed in Table I.

The following parameters were considered for evaluation of the proposed method: [16]

- Selection accuracy (SA)
- False alarm rate (FAR)
- Success rate (SR)
- Selection time (ST)

These parameters are calculated as follows: [17]

$$SA = \frac{TP}{TP + FN} X100\% \quad (1)$$

TABLE I. LIST OF TASKS PERFOEMED BY USERS

| No. of Task | Name of Task |
|---|---|
| 1 | Navigation through files and directories |
| 2 | Copy a block of text |
| 3 | Paste a block of text |
| 4 | Edit a block of text |
| 5 | Cut a block of text |
| 6 | Copy a selected objects |
| 7 | Paste a selected objects |
| 8 | Cut a selected objects |
| 9 | Undo/Redo |
| 10 | Type or delete a word |

$$FAR = \frac{FP}{TP + FP} X100\% \quad (2)$$

$$SR = \frac{SA}{SA + FAR} X100\% \quad (3)$$

where, TP= Correct Selections Performed, FP=Incorrect Selections Performed, FN=Missed Objects. Selection time (ST) is calculated from the data obtained from users.

*a) System Performance Evaluation:* The comprehensive performance of the entire system primarily depends on the scanning time of the system. We have conducted an experiment to test how the system performs in different scanning time ranging from 0.5s to 1.0s. The system performance in different scanning time is shown in Table II. Figure 12 shows a graphical representation of the average selection time, average selection, average false alarm rate, average success rate with respect to scanning time.

It is noticeable from the graph that average selection time increase along with scanning time. Average selection accuracy differs from 64% to 87%, false alarm rate ranges from 13.1% with the highest success rate in 1.0s to 2.7% with the lowest success rate in scanning time 0.5s. Increasing scanning time minimize the average FAR and increase the SR but it decreases the speed of the system. Therefore, the system needs to adjust a suitable scanning time to minimize the FAR and improve selection accuracy.

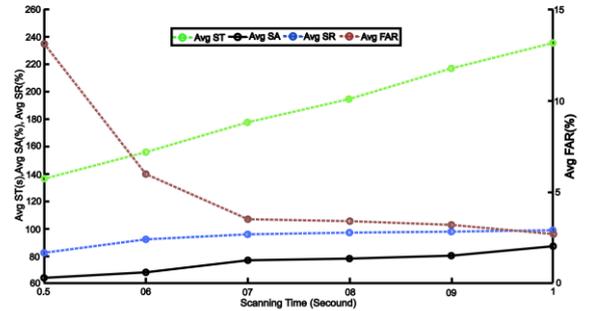

Fig. 12 The performance of proposed system in different scanning time

TABLE II. SYSTEM PERFORMANCE USING DIFFERENT SCANNING TIME

| Scanning Time | Average Selection Time | Average SA (%) | Average FAR (%) | Average SR (%) |
|---|---|---|---|---|
| 0.5 | 136.3 | 64 | 13.1 | 82.3 |
| 0.6 | 155.5 | 68 | 6.0 | 92.1 |
| 0.7 | 177.2 | 77 | 3.5 | 96 |
| 0.8 | 194.7 | 78 | 3.4 | 96.5 |
| 0.9 | 217.1 | 80 | 3.2 | 97.3 |
| 1.0 | 235.5 | 87 | 2.7 | 98.1 |

*b) Performance of Users:* The proposed system has tested on 12 users. The users were instructed to perform 10 tasks listed in table I. The performance of the user 1 and the required time for an individual task to perform with scanning time 1.0s are presented in Table III. The consolidated data of 12 users have shown in table IV. The users need to practice to operate the system using blinks effectively. Gradually, it will take less time to operate the system successfully.

The data presents true positive (TP) i.e, correct selection performed, false positive (FP) i.e, incorrect selection, false negative (FN) i.e, missed selections. Moreover, percentage of selection accuracy (SA), false alarm rate (FAR) and success rate (SR) are being calculated using equation (1), (2), and (3). The average values of selection time, selection accuracy, false alarm rate and success rate is 235.5, 87%, 2.7%, and 98.1%, respectively.

TABLE III. PERFORMANCE OF USER 1 USING SCANNING TIME=1S

| Task No. | Completion Status | Time required (s) | Wrong Selection |
|---|---|---|---|
| 1 | Yes | 300 | No |
| 2 | Yes | 240 | No |
| 3 | Yes | 120 | No |
| 4 | No | 360 | Yes |
| 5 | Yes | 540 | No |
| 6 | No | 210 | Yes |
| 7 | Yes | 80 | No |
| 8 | Yes | 380 | No |
| 9 | Yes | 65 | No |
| 10 | Yes | 60 | No |

TABLE IV. PERFORMANCE OF USERS USING PROPOSED METHOD (CONSOLIDATED DATA OF 10 USERS)

| User | No. of Task Performed | TP | FP | FN | SA (%) | FAR (%) | SR (%) |
|---|---|---|---|---|---|---|---|
| 1 | 10 | 8 | 1 | 2 | 80 | 11 | 89 |
| 2 | 10 | 9 | 0 | 1 | 90 | 0 | 100 |
| 3 | 10 | 9 | 0 | 1 | 90 | 0 | 100 |
| 4 | 10 | 8 | 0 | 2 | 80 | 0 | 100 |
| 5 | 10 | 8 | 1 | 2 | 80 | 11 | 87.9 |
| 6 | 10 | 9 | 0 | 1 | 90 | 0 | 100 |
| 7 | 10 | 10 | 0 | 0 | 100 | 0 | 100 |
| 8 | 10 | 9 | 1 | 1 | 90 | 10 | 100 |
| 9 | 10 | 8 | 0 | 2 | 80 | 0 | 100 |
| 10 | 10 | 8 | 0 | 1 | 80 | 0 | 100 |
| 11 | 10 | 9 | 0 | 1 | 90 | 0 | 100 |
| 12 | 10 | 10 | 0 | 0 | 100 | 0 | 100 |

*b) Comparison with existing system:* We have compared the proposed system with the existing systems. The comparison is listed in Table V. Moreover, the pictorial representation of users review to measure easiness is sown in fig. 13.

We have collected the review of 12 users. 8 users noted the system as manageable, 3 of them reported that easy, 1 noted as hard and no one reported as very hard or very easy.

TABLE V. COMPARISON WITH EXISTING SYSTEMS

| Reference Paper | Comparison | | |
|---|---|---|---|
| | *Lightweight* | *Easy to use* | *Harmless* |
| [3] Need a webcam and assistance to others to set the camera in perfect place | Yes | No | Yes |
| [6] Used a headset and need head and eyebrow movement, so possibilities of pain in muscles' of shoulder, head and forehead. | No | No | No |
| [7] Wear a Bluetooth headset with a sensor on forehead and three other sensor in left ear. Hence, possibilities of pain in muscles' of shoulder, head and forehead. | No | No | No |
| [8.9, 10] Need a medium quality webcam to detect face and eye position, therefore, need assistance of others to set the camera in perfect place. | Yes | No | Yes |
| The proposed system used one IR sensor with glass (no physically attached device), require no assistance of others. | Yes | Yes | Yes |

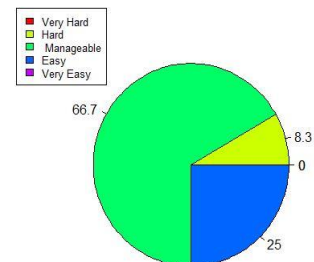

Fig. 13. Review of the users on Easiness

## V. Conclusion

An assistive approach for HCI system based on block scanning screen objects using eye blinks has been discussed in this paper. The performance of the proposed system has been evaluated by selection time, selection accuracy, false alarm rate and average success rate. The user performances have been tested through performing 10 selected tasks with scanning time 1.0s. Elderly and people with motor impairments can use this system without receiving help of others to operate the computer. The system has not implemented any camera. The blinks is captures by IR sensor, transferred by Bluetooth module, HID and ADC system of Arduino Pro micro converts analog signal to digital signal. Thus operations on monitor are being controlled without conventional mouse. In the developed user interface, after selecting an object by switching and scanning screen objects copy, cut, paste options appear and can be selected using blink while in conventional mouse controlled system they are performed using right click.